\documentclass[journal, onecolumn, a4paper]{IEEEtran}

\usepackage[utf8]{inputenc}
\usepackage{mathtools}
\usepackage{xcolor}
\usepackage{amssymb}
\usepackage{amsthm}
\usepackage{amsxtra}
\usepackage{amsmath}
\usepackage[artemisia]{textgreek}
\usepackage{array}
\usepackage{algpseudocode}
\usepackage{algorithm}
\usepackage{times}

\usepackage{verbatim}
\usepackage{epsfig}
\usepackage{setspace}
\usepackage{caption}
\usepackage{subcaption}
\usepackage{breqn}
\usepackage{color}
\usepackage{enumerate}
\usepackage{graphicx}
\usepackage{cite}
\usepackage{microtype}

\usepackage{tikz}
\usetikzlibrary{shapes, arrows.meta, positioning}

\graphicspath{{Figures/}}

\newcommand{\ex}{\mathbb{E}}
\newcommand{\pr}{\Pr}

\newcommand{\tpi}{\tilde{\pi}}

\newcommand{\BSC}{\text{BSC}}
\newcommand{\BEC}{\text{BEC}}

\newcommand{\cc}{\text{CC}}
\newcommand{\tm}{\tilde{m}}
\newcommand{\pe}{\text{P}_{\text{e}}}
\newcommand{\capint}{\text{C}_{\text{I}}}
\newcommand{\capsh}{\text{C}_{\text{Sh}}}
\newcommand{\assign}{\leftarrow}
\newcommand{\msg}{\mathsf{msg}}
\newcommand{\E}{\mathbb{E}}
\newcommand{\hite}{\text{Hit}_\text{tr}}

\algdef{SE}[SUBALG]{Indent}{EndIndent}{}{\algorithmicend }%
\algtext*{Indent}
\algtext*{EndIndent}

\newtheorem{theorem}{Theorem}
\newtheorem{definition}{Definition}
\newtheorem{lemma}{Lemma}

\newtheorem{corollary}[theorem]{Corollary}

\newtheorem*{remark*}{Remarks}
\allowdisplaybreaks

\title{Improved bounds on the interactive capacity \\ via error pattern analysis}

\author{
\IEEEauthorblockN{Mudit Aggarwal$^\dag$} \hspace{1cm} \and \IEEEauthorblockN{Manuj Mukherjee$^\ddag$}
}

\begin{document}

\maketitle
\renewcommand{\thefootnote}{}
\footnotetext{

\noindent $^\dag$M.\ Aggarwal is with the Department of Mathematics, University of British Columbia, Vancouver (UBC), British Columbia, Canada. Email: muditagg@math.ubc.ca\\

\noindent $^\ddag$M.\ Mukherjee is with the Department of Electronics and Communications Engineering, Indraprastha Institute of Information Technology Delhi (IIIT D), New Delhi, India. Email: manuj@iiitd.ac.in.
}
\renewcommand{\thefootnote}{\arabic{footnote}}

\begin{abstract}
    Any interactive protocol between a pair of parties can be reliably simulated in the presence of noise with a multiplicative overhead on the number of rounds (Schulman 1996). The reciprocal of the best (least) overhead is called the \emph{interactive capacity} of the noisy channel. 
    
    In this work, we present lower bounds on the interactive capacity of the binary erasure channel. Our lower bound improves the best-known bound due to Ben-Yishai et al. 2021 by roughly a factor of 1.75. The improvement is due to a tighter analysis of the correctness of the simulation protocol using \emph{error pattern analysis}. More precisely, instead of using the well-known technique of bounding the least number of erasures needed to make the simulation fail, we identify and bound the probability of specific erasure patterns causing simulation failure. 
    
    We remark that error pattern analysis can be useful in solving other problems involving stochastic noise, such as bounding the interactive capacity of different channels.
\end{abstract}

\section{Introduction}\label{sec:intro}

Consider a pair of parties, Alice and Bob, connected by bidirectional links, and running an \emph{alternating}\footnote{A protocol is said to be alternating if Alice and Bob speak alternately.} interactive protocol. We are interested in the scenario where the links are replaced by noisy channels. In a seminal work, Schulman \cite{S96} showed any protocol can be reliably simulated over noisy channels with a multiplicative overhead on the number of rounds of the protocol. The reciprocal of the least possible multiplicative overhead for a given noisy channel is referred to as its \emph{interactive capacity}.

Kol and Raz \cite{KR13} initiated the study of interactive capacity by showing that the binary symmetric channel with crossover probability $\epsilon$ ($\BSC(\epsilon)$) has an interactive capacity of $1-\Theta(\sqrt{h(\epsilon)})$, where $h(\cdot)$ denotes the binary entropy function. The interactive capacity of $\BSC(\epsilon)$ thus has an asymptotic (as $\epsilon\to 0$) behaviour different from its Shannon capacity $1-h(\epsilon)$ \cite[Section~7.1.4]{CT06}. Later, Ben-Yishai et al. \cite{BKOS21} showed that the interactive capacity scales linearly with the Shannon capacity for the class of \emph{binary memoryless symmetric} channels, which includes the binary symmetric and the binary erasure channels. A non-asymptotic characterization of the interactive capacity for any discrete memoryless channel remains an open problem. 

\subsection{Results}\label{sec:results}

In this work, we focus on the binary erasure channel with erasure probability $\epsilon$ ($\BEC(\epsilon)$) and obtain lower bounds on its interactive capacity. Specifically, we show the following.

\begin{theorem}
    A binary erasure channel with erasure probability $\epsilon$ satisfies
    $$
    \capint(\epsilon)\geq 0.104\:\capsh(\epsilon),
    $$
    where $\capint(\epsilon)$ and $\capsh(\epsilon)$ respectively denote the interactive and Shannon capacities of $\BEC(\epsilon)$.
    \label{th:main}
\end{theorem}
The bound in Theorem~\ref{th:main} improves the best known bound of $\capint(\epsilon)\geq 0.0604\:\capsh(\epsilon)$ due to Ben-Yishai et al. \cite[Theorem~1]{BKOS21} by roughly a factor of 1.75. 

We note here that Theorem~1 of \cite{BKOS21} actually states the constant in the bound as 0.0302. This reduction by half results from the fact that they define interactive capacity using \emph{non-adaptive protocols}\footnote{A protocol is said to be nonadaptive if the order of speaking of Alice and Bob is fixed beforehand.} with an arbitrary order of speaking. Before simulation, the authors of \cite{BKOS21} convert arbitrary nonadaptive protocols to alternating protocols by increasing the number of exchanges by at most a factor of two, thereby halving the constant in the bound. Likewise, following the same strategy, our Theorem~\ref{th:main} will yield a constant 0.052, if we were to define interactive capacity using arbitrary, and not alternating, order of speaking. In other words, irrespective of the definition used, we maintain the exact same improvement over the bound of \cite{BKOS21}.

\subsection{Techniques}\label{sec:technique}

We bound the interactive capacity of $\BEC(\epsilon)$ by constructing an explicit simulation protocol. The protocol is inspired by the protocol for adversarial erasures due to Efremenko, Gelles, and Haeupler \cite{EGH16}. However, our protocol improves theirs by reducing the size of the protocol alphabet from six to four.\footnote{We note that \cite{EGH16} also offers a different 4-symbol protocol for the erasure channel, but it is quite different from the one we design.} This reduction is crucial as we are explicitly interested in bounding the multiplicative overhead in the number of rounds. 

Our key contribution, however, is in the way we analyze the progress of the protocol, which we refer to as \emph{error pattern analysis}, and we believe it is of independent interest. In more detail, we first note that the standard recipe for analyzing protocols is defining a \emph{potential function}, and showing that the simulation succeeds if the potential function exceeds some value. Next, a \emph{worst-case} bound on the number of errors is obtained which ensures that the potential function fails to exceed the desired value. Then, the analysis is completed by evaluating the probability that the number of errors exceeds this worst-case bound via Chernoff bounds --- see for example \cite{S96,RS94,BKOS21}. We remark that while a similar analysis of our protocol is possible, this analysis is not tight, and leads to poor bounds on the interactive capacity. In fact, the bound obtained using the above technique fails to beat the pre-existing bound by \cite{BKOS21}.

We point out that the key weakness of the above technique is the fact that we analyze based on the worst-case (minimum) number of errors that is sufficient to corrupt the simulation. While such an analysis is tight for adversarial errors, it is not so for stochastic errors. For example, in our protocol, if half of the transmissions are erased, the simulation will fail if alternate transmissions were erased, but will succeed if the first half of the transmissions were erasure-free. To tighten the analysis, one thus needs to bound the probability of only those erasure patterns which will lead to simulation failure.

To do so, we model the round-by-round progress of the potential function of the protocol by a \emph{Markov process with rewards} \cite{howard}. The state transition probability of this Markov chain is determined by the channel noise, and the reward associated with a state transition is the amount by which the potential function is increased on making the said transition. The analysis then bounds the probability that the sum of the collected rewards, i.e., the value of the potential function, fails to exceed the amount necessary to guarantee the success of the simulation. This is done by first computing the expected rewards using a recurrence relation \cite[Chapter~2]{howard} and then evaluating it \cite[Chapter~8]{bona}. The analysis is then completed by bounding the probability of deviation of the collected rewards from its expected value, using a concentration inequality by Moulos \cite{moulos20} developed in the context of multi-armed bandits.

\subsection{Related works}\label{sec:rel}

Following Schulman's pioneering work on simulating interactive protocols over noisy channels, several variants of the problem have been studied --- see \cite{gelles17} for a survey. One specific direction has been to consider adversarial noise, as in \cite{S96,BR14,EGH16,GZ23,GH17}, and investigate what fraction of adversarial errors can be tolerated. Simulation of multiparty interactive protocols has also been studied in the presence of stochastic noise \cite{RS94,JSAIT,EKS18,EKPS21,clique,cycle,Star}, adversarial noise \cite{JKL15,HS16}, insertion-deletion noise \cite{GKR21,GKR22}, and beeping models \cite{EKSbeep,AGL22}.

On the other hand, following Kol and Raz's work which designed randomised protocols achieving the interactive capacity of $\BSC(\epsilon)$ asymptotically, Gelles et al. \cite{GHKRW} designed deterministic simulation protocols achieving the same. Interactive capacity has also been studied in the context of adversarial errors. Haeupler \cite{Haeupler14} investigated interactive capacity in presence of adversarial errors by allowing adaptive\footnote{Adaptive means that the simulation protocol doesn't have a predetermined order of speaking, and is instead dependent on the messages exchanged so far.} simulation, and showed that adaptive simulation increases the lower bound on interactive capacity from $1-O(\sqrt{\epsilon\log\frac{1}{\epsilon}})$ (due to Kol and Raz \cite{KR13}) to $1-O(\sqrt{\epsilon\log\log\frac{1}{\epsilon}})$. A matching upper bound is unfortunately unknown. Without using the power of adaptive simulations, Cohen and Samocha \cite{CS20} designed deterministic protocols using \emph{palette alternating tree codes} that achieve the interactive capacity of $1-\Theta(\sqrt{h(\epsilon)})$ for adversarial bit flips. Haeupler \cite{Haeupler14} also showed that using adaptive simulation, the interactive capacity of $\BSC(\epsilon)$ can be improved to at least $1-O(\sqrt{\epsilon})$. Again, a matching lower bound is missing. 

\subsection{Notation and organization}\label{sec:org}

For any string $x$, we use $x_i$ to denote the $i$th letter of $x$, and $x^n$ denotes the prefix $(x_1,x_2,\ldots,x_n)$. The length of the string $x$ is denoted by $|x|$. Given any bit $b$, its complement is denoted by $\bar{b}$. We use $\mathbf{1}\{\cdot\}$ to denote the indicator function. 

The paper is organized as follows. The setup is formally described in Section~\ref{sec:prelim} and a stronger version of Theorem~\ref{th:main} is stated. The simulation protocol which leads to the lower bound on the interactive capacity is defined in Section~\ref{sec:alg}. The correctness of the simulation protocol is proved in Section~\ref{sec:proof}. In particular, Section~\ref{sec:potential} describes the progress of the simulation protocol, and Section~\ref{sec:errorpattern} uses error pattern analysis to bound the probability of error of the simulation protocol. The proof of Theorem~\ref{th:main} appears in Section~\ref{sec:repeat}. The paper concludes with Section~\ref{sec:conc}.

\section{Problem description and results}\label{sec:prelim}

A protocol $\pi$ between a pair of parties Alice and Bob, with respective inputs $x_A$ and $x_B$, and connected by a bidirectional link, consists of a sequence of alternating\footnote{It is enough to consider alternating messages as any non-adaptive two-party protocol can be converted to an alternating protocol by increasing the number of transmissions at most by a factor of two.} bit exchanges. Without loss of generality, we shall assume throughout the paper that Alice transmits in the odd rounds, and Bob transmits in the even round. 

The transmissions are \emph{interactive}. In more detail, suppose we are in an odd round $r$, and let $m^{r-1}\in\{0,1\}^{r-1}$ be the string of $r-1$ bits sent and received by Alice in the previous rounds. Notice that in the absence of noise $m^{r-1}$ is known exactly at Bob. Then, Alice sends the bit $m_r=\pi(x_A,m^{r-1})$ to Bob in the $r$th round. We say the protocol $\pi$ has a \emph{communication complexity} $\cc(\pi)=n_0$, if there were a total of $n_0$ transmissions in $\pi$. The output of $\pi$ is the string of exchanged bits $m^{n_0}$ which is referred to as its \emph{transcript}.

In this work, we are interested in the situation where the bit exchanges between Alice and Bob are noisy. We assume that any bit sent by Alice to Bob (or vice-versa) passes through a binary erasure channel of erasure probability $\epsilon\in(0,1)$, denoted by $\BEC(\epsilon)$. The goal of Alice and Bob is now to recover the transcript $m^{n_0}$ of the original protocol $\pi$ with high probability. 

To do so, Alice and Bob are now allowed to run a new interactive protocol $\tpi$, referred to as the \emph{simulation} of $\pi$ over $\BEC(\epsilon)$. We allow the simulation protocol to have a non-alternating order of speaking, but it must be \emph{non-adaptive}, i.e., the order of speaking is fixed beforehand. In the simulation protocol, the messages sent by Alice (or Bob) in any given round now depend not only on their respective inputs and received/sent messages so far, but also on the original protocol $\pi$. Let $\cc(\tpi)=n$, and let $m_A^n$ and $m_B^n$ denote the random strings corresponding to the respective transcripts of Alice and Bob obtained by running $\tpi$ in the presence of erasures. Alice (and likewise Bob) then return estimates $\tm_A^{n_0}$ (resp. $\tm_B^{n_0}$) of the transcript $m^{n_0}$ of the original protocol $\pi$, based on $\pi$, their inputs $x_A$ (resp., $x_B$) and their transcripts $m_A^n$ (resp., $m_B^n$). We say that the simulation $\tpi$ incurred an error if either $\tm_A^{n_0}\neq m^{n_0}$ or $\tm_B^{n_0}\neq m^{n_0}$. We denote by $\pe(\tpi)$ the probability of error of $\tpi$.

Next, we define an achievable rate of interactive coding for $\BEC(\epsilon)$ as follows.
\begin{definition}
    Any $R\geq 0$ is said to be an \emph{achievable interactive coding rate} for $\BEC(\epsilon)$, if for any sequence of interactive protocols $\pi_{n_0}, n_0\geq 1,$ with $\cc(\pi_{n_0})=n_0$, there exists a corresponding sequence of simulations $\tpi_{n_0}$ over the $\BEC(\epsilon)$ satisfying $\frac{n_0}{\cc(\tpi_{n_0})}\to R$ and $\pe(\tpi_{n_0})\to 0$ as $n_0\to\infty$.
    \label{def:rate}
\end{definition}
The \emph{interactive capacity} of the $\BEC(\epsilon)$, denoted by $\capint(\epsilon)$, is defined as the supremum of all achievable interactive coding rates for $\BEC(\epsilon)$. 

It is easy to see that $\capint(\epsilon)\leq\capsh(\epsilon)$, where $\capsh(\epsilon)$ denotes the \emph{Shannon capacity} of $\BEC(\epsilon)$ --- for example, see the paragraph following equation (3) of \cite{BKOS21}. 

Recall that our main result is an improved lower bound on $\capint(\epsilon)$, stated in Theorem~\ref{th:main} in Section~\ref{sec:results}. We shall prove Theorem~\ref{th:main} by proving the following stronger result.
\begin{theorem}
    A binary erasure channel with erasure probability $\epsilon\in(0,1)$ satisfies
    $$
    \capint(\epsilon)\geq \frac{(1-\epsilon)^2}{2(2-(1-\epsilon)^2)}.
    $$
    \label{th:strong}
\end{theorem}

The subsequent sections prove Theorem~\ref{th:strong} and derive Theorem~\ref{th:main} from Theorem~\ref{th:strong}.

\section{A simulation protocol for the erasure channel}\label{sec:alg}

Fix any protocol $\pi$ with $\cc(\pi)=n_0$. In this section, we describe the simulation $\tpi$, with communication complexity $\cc(\tpi)=2kn_0$, for some $k>1$ specified later, which will be used to obtain the lower bound of Theorem~\ref{th:strong}. We note that the simulation $\tpi$ is a strengthening of Algorithm~3 of \cite{EGH16}, obtained by reducing the size of the protocol alphabet from six to four.

\begin{algorithm}[t]
\caption{The simulation $\tpi$ at party $P\in\{A,B\}$}
\begin{algorithmic}[1]
\Statex\textbf{Input:} 
a. Original protocol $\pi$ with $\cc(\pi)=n_0$; 
\Statex \hspace{1.07cm}b. Constant $k>1$ to be specified later.
\Statex
\Statex \textbf{Initialize:} $\tm_A,\tm_B\assign\emptyset$, $p_A\assign 0$, $p_B\assign 1$
\Statex
\For{$i=1$ to $n=kn_0$}
\If{sender}
\If{fresh update is possible according to $\tm_P$}
\State $t_P\assign \pi(x_P,\tm_P)$
\State $p_P\assign \bar{p}_P$
\State $\tm_P\assign\tm_P\circ t_P$ \Comment{Transcript update}
\State $\msg_P\assign (t_P,p_P)$
\State Send $\msg_P$ and store it in memory
\Else
\State Send $\msg_P$ from memory
\EndIf
\Statex
\Else
\State Receive $(t',p')$
\If{$(t',p')$ has no erasures}
\If{$P=A$ and $p'=\bar{p}_A$}
\State $\tm_A\assign\tm_A\circ t'$ \Comment{Transcript update}
\EndIf
\If{$P=B$ and $p'=p_B$}
\State $\tm_B\assign\tm_B\circ t'$ \Comment{Transcript update}
\EndIf
\EndIf
\EndIf
\EndFor
\Statex
\Statex
\State \Return $\tm_P$ truncated to length $n_0$
\end{algorithmic}
\label{alg:sim}
\end{algorithm}

In more detail, the protocol $\tpi$ consists of $kn_0$ alternating rounds, where in odd (resp., even) rounds, Alice (resp. Bob) sends a pair of bits $(t_A,p_A)$ (resp., $(t_B,p_B)$). The bit $t_A$ is a message bit, and the bit $p_A$ is a parity-check bit. These bits are computed as follows. Initially, $p_A$ is set to 0 and $p_B$ is set to 1. Both parties initialize their respective estimated transcript $\tm_A,\tm_B$, to the empty string, and update them in a step-by-step fashion as detailed below. 

Consider an odd round, that is when Alice is a sender. If $\mid\tm_A\mid$ is even, meaning Alice has to send the transmission according to $\tm_A$ and $\pi$, Alice computes a fresh message $t_A=\pi(x_A,\tm_A)$, and flips parity bit $p_A$, and sends them to Bob.\footnote{If $\mid\tm_A\mid$ exceeds or equals $n_0$, i.e., all the transmissions of $\pi$ are exhausted, simply set $t_A=0$.} Also, Alice adds $t_A$ to its estimated transcript $\tm_A$. If on the other hand, if $\mid\tm_A\mid$ was odd, Alice needs the next bit from Bob to compute its next transmission. In such a scenario, Alice simply sends the pair of bits it had sent in the previous odd round. Furthermore, no update is made to $\tm_A$. Bob follows a similar set of rules on even rounds. 

In the receiver mode, that is Alice in even rounds and Bob in odd rounds, if both received bits are non-erased, the parties decide to update their estimated transcripts based on the received parity. The detailed scheme appears in Algorithm~\ref{alg:sim}.

\section{Proof of Theorem~\ref{th:strong}}\label{sec:proof}

In this section, we prove Theorem~\ref{th:strong}. The idea is to show that the simulation $\tpi$ in Algorithm~\ref{alg:sim} achieves a rate of $\frac{(1-\epsilon)^2}{2(2-(1-\epsilon)^2)}$, while ensuring $\pe(\tpi)\to 0$ as $n_0\to\infty$. The proof is divided into two parts. Section~\ref{sec:potential} analyzes how erasures affect the updating of the estimated transcripts in Algorithm~\ref{alg:sim}, and concludes by identifying a condition for correctness of simulation based on the lengths of the returned estimated transcripts. Section~\ref{sec:errorpattern} then introduces our key contribution, the error-pattern analysis, using which we bound the probability of occurrence of the specific error patterns that lead to the violation of the correctness condition identified in Section~\ref{sec:potential}. 

\subsection{Progress analysis of Algorithm~\ref{alg:sim}}\label{sec:potential}

We denote by $\tm_P^{(i)}$ and $p_P^{(i)}$ for $P\in\{A,B\}$, the estimated transcript and the parity bit of party $P$, respectively, \emph{at the beginning} of the $i$th round of $\tpi$. This section will prove the following result which relates the correctness of the simulation $\tpi$ with the the lengths of the estimated transcripts.

\begin{theorem}
    The simulation $\tpi$ returns the correct transcripts if $\vert\tm_A^{(kn_0+1)}\vert+\vert\tm_B^{(kn_0+1)}\vert\geq 2n_0$.
    \label{th:correct}
\end{theorem}

To proceed, note that $\mid\tm_P^{(i)}\mid, P\in\{A,B\}$ is non-decreasing in $i$. Next, observe that $p_A$ is flipped if and only if $\mid\tm_A\mid$ is increased from even to odd. Since, initially $p_A$ was assigned 0, we must have $p_A^{(i)}$ being one if and only if $\mid\tm_A^{(i)}\mid\mod 4\in\{2,3\}$. Similarly, we can argue that $p_B^{(i)}$ is one if and only if $\mid\tm_B^{(i)}\mid\mod 4\in\{1,2\}$. For the sake of brevity, we define $s_P^{(i)}\triangleq\mid\tm_P^{(i)}\mid\mod 4$ for $P\in\{A,B\}$. Table~\ref{tab:status} lists down the relationship between the lengths of the estimated transcripts and parity.

\begin{table}
    \centering
    \begin{tabular}{|c|c|c|}
    \hline
    $s_P^{(i)}$ & $p_A^{(i)}$ & $p_B^{(i)}$ \\

    \hline
    0 & 0 & 1\\
    \hline
    1 & 1 & 1\\
    \hline
    2 & 1 & 0\\
    \hline
    3 & 0 & 0\\
    \hline
    \end{tabular}
    \caption{Relationship between $s_P^{(i)}$ and $p_P^{(i)}$}
    \label{tab:status}
\end{table}

We now prove a series of technical lemmas to conclude that $\tpi$ leads to correct simulation if $\mid\tm_A^{(kn_0+1)}\mid+\mid\tm_B^{(kn_0+1)}\mid\geq 2n_0$. 

\begin{lemma}
    For any $1\leq i\leq kn_0$, $\biggl\vert\mid\tm_A^{(i)}\mid-\mid\tm_B^{(i)}\mid\biggr\vert\leq 1$ hold.
    \label{lem:gap}
\end{lemma}

\begin{IEEEproof}
    We shall do this by induction on $i$. By the initialization to null strings, the base case of $i=1$ is obvious. So we assume that the condition holds up to the beginning of the $i$th round. We shall show that following round $i$, we have $\biggl\vert\mid\tm_A^{(i+1)}\mid-\mid\tm_B^{(i+1)}\mid\biggr\vert\leq 1$.

    Firstly, note that in any given round, the gap between the lengths of a pair of estimated transcripts cannot increase by more than one. Hence, the induction hypothesis at $i+1$ is obvious if $\mid\tm_A^{(i)}\mid=\mid\tm_B^{(i)}\mid$ was true. We now prove that the induction hypothesis still holds at $i+1$ if $\mid\tm_A^{(i)}\vert=\vert\tm_B^{(i)}\vert+1$. The remaining case when $\mid\tm_B^{(i)}\vert=\vert\tm_A^{(i)}\vert+1$ can be argued similarly.

    We begin by showing that $\mid\tm_A^{(i)}\vert=\vert\tm_B^{(i)}\vert+1$ implies that $\mid\tm_A^{(i)}\mid$ is odd. Suppose it is not. Let $j\leq i-1$ be the round in which $\tm_A^{(i)}$ reached its current length, that is, $\vert\tm_A^{(j)}\vert+1=\vert\tm_A^{(i)}\vert$ and $\vert\tm_A^{(j+1)}\vert=\vert\tm_A^{(i)}\vert$. Now, if $\mid\tm_A^{(i)}\mid$ is even, then $j$ must have been an even round in which Bob was the sender and Alice the receiver. Since, Alice accepted the received bit in round $j$, we must have $p_A^{(j)}=p_A^{(j+1)}=\bar{p}_B^{(j+1)}$. Next, note using the non-decreasing property of transcript length that $\mid\tm_B^{(j+1)}\vert\leq\vert\tm_B^{(i)}\vert<\vert\tm_A^{(i)}\vert=\vert\tm_A^{(j+1)}\vert$. Then, noting that $\mid\tm_A^{(j+1)}\vert=\vert\tm_A^{(i)}\vert$ was assumed to be even, the induction hypothesis at $j+1$ and $\mid\tm_B^{(j+1)}\vert<\vert\tm_A^{(j+1)}\vert$ ensures that no entry in Table~\ref{tab:status} is consistent with $p_A^{(j+1)}=\bar{p}_B^{(j+1)}$. Hence, we have a contradiction, and our assumption $\mid\tm_A^{(i)}\mid$ being even was wrong.

    Now, assume $i$ was odd. With $\mid\tm_A^{(i)}\mid$ being odd, Alice cannot be sending a fresh message. Hence, $\mid\tm_A\mid$ doesn't increase in length, and irrespective of $\mid\tm_B\mid$ increasing its length, the induction hypothesis continues to hold at $i+1$. It remains to prove the case when $i$ is even.

    When $i$ is even, if $\mid\tm_B\mid$ increases in round $i$, the induction hypothesis continues to hold for $i+1$. We therefore deal with the case that $\mid\tm_B\mid$ did not increase in round $i$ by showing that $\mid\tm_A\mid$ also did not increase. Now $\mid\tm_B\mid$ not increasing occurs only if $\mid\tm_B^{(i)}\mid$ is even. Hence, either $(s_B^{(i)}=0,s_A^{(i)}=1)$ or $(s_B^{(i)}=2,s_A^{(i)}=3)$ as $\mid\tm_A^{(i)}\mid=\mid\tm_B^{(i)}\mid+1$. Therefore, by Table~\ref{tab:status}, we have, either $(p_A^{(i)}=1,p_B^{(i)}=1)$ or $(p_A^{(i)}=0,p_B^{(i)}=0)$, respectively. Now, since $\mid\tm_B\mid$ was not increased, we must have $p_B^{(i)}=p_B^{(i+1)}$. Also, $p_A^{(i+1)}=p_A^{(i)}$, since Alice was a receiver in round $i$. Therefore, we must have either $p_A^{(i+1)}=0,p_B^{(i+1)}=0$ or $p_A^{(i+1)}=p_B^{(i+1)}=1$, and Alice doesn't accept the received bit in both cases. Hence, $\mid\tm_A\mid$ also doesn't change in round $i$ as required. This completes the proof of the induction hypothesis.
\end{IEEEproof}

We note down an important corollary of Lemma~\ref{lem:gap} which we have proved in the course of proving Lemma~\ref{lem:gap}.

\begin{corollary}
    For any $1\leq i\leq kn_0$, we have that $\mid\tm_A^{(i)}\mid\neq\mid\tm_B^{(i)}\mid$ implies $\mid\tm_A^{(i)}\mid$ is odd and $\mid\tm_B^{(i)}\mid$ is even.
    \label{cor:oddeven}
\end{corollary}

\begin{IEEEproof}
    Note that we proved $\mid\tm_A^{(i)}\mid=\mid\tm_B^{(i)}\mid+1$ implies $\mid\tm_A^{(i)}\mid$ is odd, while proving Lemma~\ref{lem:gap}. Similarly, we can prove $\mid\tm_B^{(i)}\mid=\mid\tm_A^{(i)}\mid+1$ implies $\mid\tm_B^{(i)}\mid$ is even. Together with Lemma~\ref{lem:gap}, they imply the corollary. 
\end{IEEEproof}

The following technical lemma describes some conditions on the equality of the lengths of the two estimated transcripts.

\begin{lemma}
    Let $\mid\tm_A^{(i)}\vert=\vert\tm_B^{(i)}\vert$. Then the following hold:
    \begin{enumerate}[i)]
        \item If $\mid\tm_A^{(i)}\mid$ is even, then $i$ is odd.
        \item If $\mid\tm_A^{(i)}\mid$ is odd, then $i$ is even.
    \end{enumerate}
    \label{lem:eq}
\end{lemma}

\begin{IEEEproof}
    We shall only prove i). The second statement ii) can be proved in the exact same manner.

    We prove i) by contradiction. So, assume $\mid\tm_A^{(i)}\vert=\vert\tm_B^{(i)}\vert$ and $\mid\tm_A^{(i)}\vert$ is even, and assume that $i$ is also even. Let $j\leq i-1$ be the round in which $\tm_A$ was updated to its current length, i.e., $\vert\tm_A^{(j)}\vert+1=\vert\tm_A^{(j+1)}\vert=\vert\tm_A^{(i)}\vert$. Then, since $\mid\tm_A^{(i)}\vert$ is even, round $j$ must have been even, and Bob was the sender. Thus, noting the assumption that $i$ was even, we must have $j\leq i-2$. Then, the round $j+1<i$ is odd, and $\mid\tm_A\mid^{(j+1)}$ is even, and hence, Alice will add a fresh message resulting in $\vert\tm_A^{(j+2)}\vert=\vert\tm_A^{(j+1)}\vert+1=\vert\tm_A^{(i)}\vert+1$. This however violates the non-decreasing property of the transcripts as $j+2\leq i$. Hence we have a contradiction, and $i$ must have been odd. 
\end{IEEEproof}

The next pair of lemmas describes how the transcripts are affected by erasures.

\begin{lemma}
    Let $\mid\tm_A^{(i)}\vert=\vert\tm_B^{(i)}\vert$. Then the following hold:
    \begin{enumerate}[i)]
        \item If there were no erasures then $\vert\tm_A^{(i+1)}\vert=\vert\tm_B^{(i+1)}\vert=\vert\tm_A^{(i)}\vert+1$.
        \item If $i$ is odd and there was an erasure, then $\vert\tm_A^{(i+1)}\vert=\vert\tm_B^{(i+1)}\vert+1=\vert\tm_B^{(i)}\vert+1$.
        \item If $i$ is even and there was an erasure, then $\vert\tm_B^{(i+1)}\vert=\vert\tm_A^{(i+1)}\vert+1=\vert\tm_A^{(i)}\vert+1$.
    \end{enumerate}
    \label{lem:erase1}
\end{lemma}

\begin{IEEEproof}
    We first prove i) when $i$ is odd. The case of $i$ even follows similarly. Firstly, note that by Lemma~\ref{lem:eq}, we have that $i$ being odd implies $\vert\tm_A^{(i)}\vert$ is even. Thus, in the $i$ round, Alice will add a fresh symbol, and hence $\vert\tm_A^{(i+1)}\vert=\vert\tm_A^{(i)}\vert+1$. If there were no erasures, then noting that $s_A^{(i)}=s_B^{(i)}\in\{0,2\}$, and $p_A^{(i+1)}=\bar{p}_A^{(i)}$ as Alice made a fresh transmission, Table~\ref{tab:status} ensures that $p_A^{(i+1)}=p_B^{(i)}$ holds. Hence, Bob will accept this bit, and so $\vert\tm_B^{(i+1)}\vert=\vert\tm_B^{(i)}\vert+1$.

    Next, we shall prove ii). Here, if $i$ is odd, by Lemma~\ref{lem:eq}, we must have that $\vert\tm_A^{(i)}\vert$ is even. Hence, Alice transmits a fresh message thereby adding a symbol to its estimated transcript. On the other hand, Bob doesn't accept the message due to the erasure, and hence its estimated transcript remains unchanged. The proof of iii) follows identically by reversing the roles of Alice and Bob.
\end{IEEEproof}

\begin{lemma}
    \begin{enumerate}[i)]
        \item If $\mid\tm_A^{(i)}\vert=\vert\tm_B^{(i)}\vert+1$, and $i$ is odd, then without any erasures we have $\mid\tm_A^{(i+1)}\vert=\vert\tm_B^{(i+1)}\vert=\vert\tm_A^{(i)}\vert$. On the other hand, if there was an erasure, then $\mid\tm_A^{(i+1)}\vert=\vert\tm_A^{(i)}\vert=\vert\tm_B^{(i+1)}\vert+1$.
        \item If $\mid\tm_A^{(i)}\vert=\vert\tm_B^{(i)}\vert+1$, and $i$ is even, then irrespective of erasures the transcripts do not change in round $i$.
        \item If $\mid\tm_B^{(i)}\vert=\vert\tm_A^{(i)}\vert+1$, and $i$ is even, then without any erasures we have $\mid\tm_B^{(i+1)}\vert=\vert\tm_A^{(i+1)}\vert=\vert\tm_B^{(i)}\vert$. On the other hand, if there was an erasure, then $\mid\tm_B^{(i+1)}\vert=\vert\tm_B^{(i)}\vert=\vert\tm_A^{(i+1)}\vert+1$.
        \item If $\mid\tm_B^{(i)}\vert=\vert\tm_A^{(i)}\vert+1$, and $i$ is odd, then irrespective of erasures the transcripts do not change in round $i$.
    \end{enumerate}
    \label{lem:erase2}
\end{lemma}

\begin{IEEEproof}
    Again, we shall only prove i) and ii), and the remaining two parts follow similarly.

    First, we prove i). Note by Corollary~\ref{cor:oddeven}, we must have $\vert\tm_A^{(i)}\vert$ to be odd, and hence Alice will not be sending a fresh message. Hence, $\vert\tm_A^{(i+1)}\vert=\vert\tm_A^{(i)}\vert$, and $p_A^{(i+1)}=p_A^{(i)}$. The case with erasures is trivial as Bob will not update his transcript. Now, assume that there were no erasures. Since, $\vert\tm_A^{(i)}\vert$ is odd, and $\mid\tm_A^{(i)}\vert=\vert\tm_B^{(i)}\vert+1$, we must have the following two possibilities, $(s_A^{(i)}=1,s_B^{(i)}=0)$ or $(s_A^{(i)}=3,s_B^{(i)}=2)$. In both cases, Table~\ref{tab:status}  ensure $p_A^{(i)}=p_B^{(i)}$. Thus, noting that $p_A^{(i+1)}=p_A^{(i)}=p_B^{(i)}$, Bob will accept the symbol, and hence $\vert\tm_B^{(i+1)}\vert=\vert\tm_B^{(i)}\vert+1$. 

    Next, we prove ii). Here, note by Corollary~\ref{cor:oddeven} that $\vert\tm_B^{(i)}\vert$ is even. Bob will thus not be sending a fresh update on even round $i$. Thus, Bob's transcript remains unchanged. If there is an erasure, Alice will not update her transcript. It therefore remains to show that Alice will not accept Bob's message in case there are no erasures. To see this, observe that $\vert\tm_B^{(i)}\vert$ is even and $\mid\tm_A^{(i)}\vert=\vert\tm_B^{(i)}\vert+1$ correspond to the cases $(s_A^{(i)}=1,s_B^{(i)}=0)$ or $(s_A^{(i)}=3,s_B^{(i)}=2)$. Since Bob doesn't send a fresh message, $p_B^{(i+1)}=p_B^{(i)}$. Now, by Table~\ref{tab:status}, $(s_A^{(i)}=1,s_B^{(i)}=0)$ or $(s_A^{(i)}=3,s_B^{(i)}=2)$ implies that $p_A^{(i)}=p_B^{(i)}=p_B^{(i+1)}$. Hence, Alice will not accept Bob's bit. 
\end{IEEEproof}

We note that by Lemma~\ref{lem:gap}, the hypotheses of Lemmas~\ref{lem:erase1} and \ref{lem:erase2}, exhausts all possible relationships between $\tm_A^{(i)}$ and $\tm_B^{(i)}$. The following lemma proves that the estimated transcripts are always prefixes of the correct transcript $m^{n_0}$ of the original protocol $\pi$.

\begin{lemma}
    The estimated transcripts $\tm_A^{(i)}$ and $\tm_B^{(i)}$ are always prefixes of the correct transcript $m^{n_0}$.
    \label{lem:prefix}
\end{lemma}

\begin{IEEEproof}
    We prove this again using induction. The base case $i=1$ is trivial, and we assume the result holds up to $i$, and show that the induction hypothesis continues to hold for $i+1$.

    To do so, first consider $\vert\tm_A^{(i)}\vert=\vert\tm_B^{(i)}\vert$. Then, using the same arguments as in the proof of Lemma~\ref{lem:erase1}, we can show that any update of the estimated transcripts is correct, as long as $\tm_A^{(i)},\tm_B^{(i)}$ are correct. Hence, by the induction hypothesis, correctness holds for $i+1$. In the same way, if $\vert\tm_A^{(i)}\vert\neq\vert\tm_B^{(i)}\vert$, we can, using exactly the same arguments as in the proof of Lemma~\ref{lem:erase2}, show that any fresh update of the transcripts in round $i+1$ are correct, as long as $\tm_A^{(i)},\tm_B^{(i)}$. The induction hypothesis at $i+1$ thus follows.
\end{IEEEproof}

We now conclude this section by proving Theorem~\ref{th:correct}

\begin{IEEEproof}[Proof of Theorem~\ref{th:correct}]
    By Lemma~\ref{lem:prefix}, it is enough to show that $\vert\tm_P^{(kn_0+1)}\vert\geq n_0$, for $P=A,B$. To see this, suppose $\vert\tm_A^{(kn_0+1)}\vert\leq n_0-1$. Hence, by Lemma~\ref{lem:gap}, we have $\vert\tm_B^{(kn_0+1)}\vert\leq n_0$. This would then violate the fact that $\vert\tm_A^{(kn_0+1)}\vert+\vert\tm_B^{(kn_0+1)}\vert\geq 2n_0$, and hence $\vert\tm_A^{(kn_0+1)}\vert\geq n_0$. Similarly, we can prove $\vert\tm_B^{(kn_0+1)}\vert\geq n_0$.
\end{IEEEproof}

\subsection{Error-pattern analysis}\label{sec:errorpattern}

In this section, we examine the random process $T_i\triangleq\vert\tm_A^{(i)}\vert+\vert\tm_B^{(i)}\vert, i\geq 1,$ and upper bound the probability that $\vert\tm_A^{(kn_0+1)}\vert+\vert\tm_B^{(kn_0+1)}\vert$ doesn't exceed $2n_0$. By Theorem~\ref{th:correct}, this translates to an upper bound on $\pe(\tpi_{n_0})$, and allows us to choose a $k$ that ensures $\pe(\tpi_{n_0})\to 0$ as $n_0\to\infty$. More precisely, we shall prove the following theorem.

\begin{theorem}
    For the choice $k>\frac{2}{(1-\epsilon)^2}-1$ in Algorithm~\ref{alg:sim}, the simulation protocol $\tpi$ satisfies $\pe(\tpi)\to 0$ as $n_0\to\infty$.
    \label{th:choice}
\end{theorem}

The remainder of this section is dedicated to proving Theorem~\ref{th:choice}. To proceed, we define $p$ to be the probability that at least one of the transmissions of a round got erased, i.e., $p=1-(1-\epsilon)^2$. Next, we define the following set of protocol states at beginning of the $i$th round: 
\begin{enumerate}[(I)]
    \item \(\mid\tm_A^{(i)}\mid = \mid\tm_B^{(i)}\mid\) and \(i\) even.
    \item \(\mid\tm_A^{(i)}\mid = \mid\tm_B^{(i)}\mid\) and \(i\) odd.
    \item \(\mid\tm_A^{(i)}\mid = \mid\tm_B^{(i)}\mid+1\) and \(i\) odd.
    \item \(\mid\tm_A^{(i)}\mid = \mid\tm_B^{(i)}\mid+1\) and \(i\) even.
    \item \(\mid\tm_A^{(i)}\mid + 1 = \mid\tm_B^{(i)}\mid\) and \(i\) odd.
    \item \(\mid\tm_A^{(i)}\mid + 1 = \mid\tm_B^{(i)}\mid\) and \(i\) even.
\end{enumerate}
Note that by Lemma~\ref{lem:gap}, the above set of states is exhaustive. 

Next, we define a Markov chain $Z_i, i\geq 1,$ with the state space $S=\{s_1,s_2,s_3\}$, where the states $s_i$ are defined as
\begin{align}
    s_1 & =\text{Protocol is in state I or II,}\notag\\
    s_2 & =\text{Protocol is in state IV or V,}\notag\\
    s_3 & =\text{Protocol is in state III or VI.}\label{eq:states}
\end{align}
The Markov chain $Z_i, i\geq 1,$ thus tracks the evolution of the protocol states round-by-round. Observe that at the beginning of the simulation $\tpi$, the protocol is in state II. Hence, $\pr(Z_1=s_1)=1$ is the initial distribution of the Markov chain. Furthermore, define the random process $R_i\triangleq T_{i+1}-T_i$, which tracks the growth of the length of the estimated transcripts over the rounds. Next, we use the lemmas in the previous section to define the state transitions. 

First, consider that $Z_i=s_1$. Then, without any erasures, i.e., with probability $1-p$, by Lemma~\ref{lem:erase1}, $Z_{i+1}=s_1$, and $R_i=2$. On the other hand, if there was erasure, Lemma~\ref{lem:erase1} tells us that protocol state I transitions to V, and state II transitions to IV. In other words, we have $Z_{i+1}=s_2$. Furthermore, we also have $R_i=1$. Next, assume $Z_i=s_2$. Here, by Lemma~\ref{lem:erase2}, state IV transitions to III and state V transitions to state VI, irrespective of whether there was an erasure or not. Thus, we will have $Z_{i+1}=s_3$, and $R_i=0$. Finally, given $Z_i=s_3$, we similarly have using Lemma~\ref{lem:erase2}, that $Z_{i+1}=s_1$ if there were no erasures, i.e., with probability $1-p$, with $R_i=1$. On the other hand, if there was an erasure, i.e., with probability $p$, Lemma~\ref{lem:erase2} shows that $Z_{i+1}=s_2$ with $R_i=0$. 

\begin{figure}[t]
    \centering
\begin{tikzpicture}[
    node distance=1.7cm,
    state/.style={circle, draw, minimum size=0.5cm},
    transition/.style={-Stealth, semithick}
]

\node[state] (A) {$s_1$};j
\node[state, right=of A] (B) {$s_2$};
\node[state, right=of B] (C) {$s_3$};

\draw[transition] (A) -- node[midway, above] {$p, \; 1$} (B);
\draw[transition] (C) to[out=220,in=320] node[midway, below] {$1-p, \; 1$} (A);
\path (A) edge [loop above] node[midway, above] {$1-p, \; 2$} (A);
\draw[transition] (B) to[out=30,in=150] node[midway, above] {$1, \; 0$} (C);
\draw[transition] (C) to[out=210,in=330] node[midway, above] {$p, \; 0$} (B);
\end{tikzpicture}
    \caption{State transition diagram of the Markov reward-on-edges process $(Z_i,R_i),i\geq 1$. The edges are labelled by the tuple (probability of transition, reward).}
    \label{fig:markov}
\end{figure}
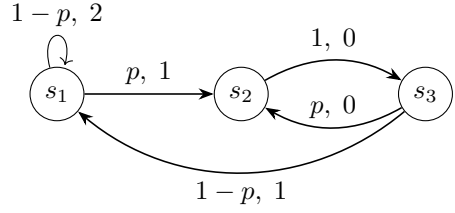

The state transition diagram of the Markov chain $Z_i,i\geq 1,$ is summarised in Figure~\ref{fig:markov}. Furthermore, note from the above discussion that $R_i$ is determined by the pair $(Z_i,Z_{i+1})$, and hence is a function of the edges of the state transition diagram. The process $(Z_i,R_i), i\geq 1,$ is known as a Markov process with rewards --- see Chapter~2 of \cite{howard}. Accordingly, we label the edges in Figure~\ref{fig:markov} by the tuple (probability of transition, reward). Note that the total reward after $i$ rounds is given by $\sum_{j=1}^{i}R_j$ which equals $T_{i+1}$ since $T_1=0$ by the initialization step of Algorithm~\ref{alg:sim}. Our goal is therefore to bound the probability of the event that $T_{kn_0+1}<2n_0$, which is equivalent to the event $\sum_{i=1}^{kn_0}R_i<2n_0$. 

To proceed, we first evaluate the expected reward after $kn_0+1$ rounds, given the chain starts ar $s_1$, i.e, $\ex[\sum_{i=1}^{kn_0}R_i\mid Z_1=s_1]$. We first note the following lemma, which is a consequence of the time-homogeneity of the Markov chain $Z_i, i\geq 1$.
\begin{lemma}
    Given any $s\in\{s_1,s_2,s_3\}$, and any $n_1,n_2\in\mathbb{N}$ with $n_1<n_2$, we have
    $$
    \ex\biggl[\sum_{i=n_1}^{n_2}R_i\mid Z_{n_1}=s\biggr] = \ex\biggl[\sum_{i=1}^{n_2-n_1}R_i\mid Z_1=s\biggr].
    $$
    \label{lem:timehomogeneity}
\end{lemma}

\begin{IEEEproof}
    The proof follows by noting the time-homogeneity of $Z_i,i\geq 1$, and the fact that $R_i$ is completely determined by the pair $(Z_i,Z_{i+1})$.
\end{IEEEproof}

Lemma~\ref{lem:timehomogeneity} therefore allows us to calculate the expectation $\ex[\sum_{i=1}^{kn_0}R_i\mid Z_1=s_1]$ via a recurrence relation.

\begin{lemma}
    Define the function $f(n)\triangleq \ex[\sum_{i=1}^{n}R_i\mid Z_1=s_1]$. Then, $f(n)$ satisfies the recurrence relation $f(n+2) = (1-p)f(n+1) + pf(n) + 2(1-p)$, with initial conditions $f(0) = 0$ and $f(1) = 2-p$.
    \label{lem:recurrence}
\end{lemma}

\begin{IEEEproof}
    We begin by defining $g(n,i)\triangleq \ex\biggl[\sum_{i=1}^n R_i\mid Z_1=s_i\biggr]$, for all $i\in\{1,2,3\}, n\geq 1$. Note that $f(n)=g(n,1)$. Now, observe that 
    \begin{align}
        g(n,1) & = \ex\biggl[\sum_{i=1}^nR_i\mid Z_0=s_1\biggr]\notag\\
             & = \ex\biggl[\ex\biggl[\sum_{i=1}^nR_i\mid Z_0=s_1,Z_1\biggr]\biggr]\notag\\
             & = \sum_{j=1}^3\pr(Z_1=s_j\mid Z_0=s_1)\ex\biggl[\sum_{i=1}^nR_i\mid Z_0=s_1,Z_1=s_j\biggr]\notag\\
             & \stackrel{(a)}{=} (1-p)\biggl(2+\ex\biggl[\sum_{i=2}^nR_i\mid Z_1=s_1,Z_0=s_1\biggr]\biggr)+p\biggl(1+\ex\biggr[\sum_{i=2}^nR_i\mid Z_1=s_1,Z_0=s_1\biggr]\biggr)\notag\\
             & \stackrel{(b)}{=} (2-p)+(1-p)\ex\biggl[\sum_{i=1}^{n-1}R_i\mid Z_0=s_1\biggr]+p\ex\biggl[\sum_{i=1}^{n-1}R_i\mid Z_0=s_2\biggr]\notag\\
             & = (2-p)+(1-p)g(n-1,1)+pg(n-1,2),\label{eq:rec:1} 
    \end{align}
    where $(a)$ follows using the transition probabilities and rewards of the process $(Z_i,R_i)$ as shown in Figure~\ref{fig:markov}, while $(b)$ uses the Markovity of $Z_i$ and Lemma~\ref{lem:timehomogeneity}. 

    Similar to \eqref{eq:rec:1}, we can also derive using Lemma~\ref{lem:timehomogeneity} the following relations:
    \begin{align}
        g(n,2) = & g(n-1,3),\label{eq:rec:2}\\
        g(n,3) = &  \; (1-p)+(1-p)g(n-1,1)+pg(n-1,2). \label{eq:ref:3}
    \end{align}
    Now, plugging in \eqref{eq:rec:2} in \eqref{eq:rec:1}, we get
\begin{align*}
    g(n+2,1) & = (2-p)+(1-p)g(n+1,1)+pg(n,3)\\
           & \stackrel{(a)}{=} (2-p)+(1-p)g(n+1,1)+p(1-p)
           +p[(1-p)g(n-1,1)+pg(n-1,2)]\\
           & \stackrel{(b)}{=} (2-p)+(1-p)g(n+1,1)+p(1-p)+pg(n,1)-p(2-p)\\
           & = 2(1-p)+(1-p)g(n+1,1)+pg(n,1),
\end{align*}
where $(a)$ and $(b)$ respectively use \eqref{eq:ref:3} and \eqref{eq:rec:1}. Recalling that $g(n,1)=f(n)$, this completes the proof of the recurrence relation $f(n+2) = (1-p)f(n+1) + pf(n) + 2(1-p)$. The initial conditions $f(0)=0$ and $f(1)=2-p$ follow by noting the rewards and the transition probabilities of the process $(Z_i,R_i)$ as shown in Figure~\ref{fig:markov}.
\end{IEEEproof}

Next, we solve the recurrence relation obtained in Lemma~\ref{lem:recurrence}. The solution is via formal power series and partial fractions --- see Chapter~8 of \cite{bona} for an overview on these methods. We recall the \emph{binomial series} in the next lemma, which will be needed to solve the recurrence.

\begin{lemma}[Binomial series]
    For any $m\in\mathbb{R}$, we have
    $$
    (1+x)^m=\sum_{n\geq 0}\binom{m}{n}x^n,
    $$
    where $\binom{m}{n}$ is the generalized binomial coefficient given by 
    $$
    \binom{m}{n}\triangleq\frac{m(m-1)\ldots(m-n+1)}{n!}.
    $$
    \label{lem:binom}
\end{lemma}
\begin{IEEEproof}
    This is a standard textbook result --- see, for example, Theorem~4.15 of \cite{bona}.
\end{IEEEproof}

\begin{lemma}
    For any $n\geq 1$, the process $(Z_i,R_i)$ satisfies $$\ex\biggl[\sum_{i=1}^nR_i\mid Z_1=s_1\biggr]=\frac{2n (1-p^2) + p(3-p)(1-(-p)^n)}{(1+p)^2}.$$
    \label{lem:expectation}
\end{lemma}

\begin{IEEEproof}
    Setting \(f(n) = \E\biggl[\sum_{i=1}^nR_i \mid Z_1 = s_1\biggr]\), we compute $f(n)$ using the recurrence relation \(f(n+2) = (1-p)f(n+1) + pf(n) + 2(1-p)\), with \(f(0) = 0\) and \(f(1) = 2-p\), as obtained in Lemma~\ref{lem:recurrence}.

Let \(F(x) = \sum_{n \geq 0}f(n) x^n\) be the generating function for \(f(n), n\geq 0\). Hence, we get:
\begin{align*}
    \sum_{n \geq 0} f(n+2) x^n = &(1-p) \sum_{n \geq 0} f(n+1) x^n + p \sum_{n \geq 0} f(n) x^n+ 2(1-p)\sum_{n \geq 0} x^n .
\end{align*}
Substituting \(F(x)\) and the values of \(f(0)\) and \(f(1)\), we get:
\begin{align*}
    \frac{F(x) - (2-p)x}{x^2} = (1-p)\frac{F(x)}{x} + pF(x) + \frac{2(1-p)}{1-x} \\
    \implies F(x) = \frac{x^2}{1-(1-p)x - px^2} \left [ \frac{2(1-p)}{1-x} + \frac{2-p}{x}\right ] \\
    \implies F(x) = x^2 \frac{2(1-p)}{(1-x)^2 (1+px)} + x \frac{2-p}{(1-x)(1+px)}
\end{align*}

We denote the coefficient of \(x^n\) in the series expansion of a function \(g(x)\) by \([x^n]g(x)\). Hence, we require \(f(n) = [x^n]F(x)\). We do that using partial fractions. 

Now, let 
\[a(x) = \frac{2(1-p)}{(1-x)^2 (1+px)} \quad \text{and}\quad b(x) = \frac{2-p}{(1-x)(1+px)}.\]

Note, for the first part:
\[\frac{1}{(1-x)^2 (1+px)} = \frac{p/(1+p)^2}{1-x} + \frac{p^2/(1+p)^2}{1+px} + \frac{1/(1+p)}{(1-x)^2};\]
and for the second part:
\[\frac{1}{(1-x) (1+px)} = \frac{1/(1+p)}{1-x} + \frac{p/(1+p)}{1+px}.\]

Hence, we get:
\begin{align*}
    f(n) = &\; [x^n]F(x) = [x^n](x^2a(x) + xb(x)) \\
    = & \;[x^{n-2}]a(x) + [x^{n-1}]b(x) \\
    = & \;2(1-p) \cdot [x^{n-2}]\left (\frac{p/(1+p)^2}{1-x} + \frac{p^2/(1+p)^2}{1+px} + \frac{1/(1+p)}{(1-x)^2} \right ) + (2-p) \cdot [x^{n-1}] \left ( \frac{1/(1+p)}{1-x} + \frac{p/(1+p)}{1+px} \right ) \\
    \stackrel{(a)}{=} &\; 2(1-p) \left ( \frac{p}{(1+p)^2} + \frac{p^2(-p)^{n-2}}{(1+p)^2} + \frac{n-1}{1+p}\right ) + (2-p)\left ( \frac{p (-p)^{n-1}}{1+p} + \frac{1}{1+p}\right ) \\
    =&\; \frac{2n (1-p^2) + p(3-p)(1-(-p)^n)}{(1+p)^2},
\end{align*}
where $(a)$ uses Lemma~\ref{lem:binom}.
\end{IEEEproof}

After having obtained $\ex[T_{kn_0+1}\mid Z_1=s_1]$, we proceed to bound $\pr(T_{kn_0+1}<2n_o)$ using a \emph{concentration inequality}. We shall use the following concentration inequality due to Moulos \cite{moulos20}. To state the inequality, we need to define the \emph{hitting time} of the \emph{transition chain} of $Z_i, i\geq 1$. The transition chain of $Z_i$ is a new Markov chain $Y_i,i\geq 1$, with the state space $S\times S$, and defined as $Y_i=(Z_{i+1},Z_i)$. Observe that the transition probabilities for the transition chain are given by $\pr(Y_{i+1}=(s,t)\mid Y_i=(s',t'))=\mathbf{1}\{t=s'\}\pr(Z_{i+1}=s\mid Z_i=t)$. Next, define the random variable $H_{s,t}\triangleq\inf\{i\geq 2: Y_i=(s,t)\}$. Then, the hitting time of the transition chain $Y_i$ is defined as $$\hite\triangleq\max_{(s,t),(s',t')\in S\times S}\ex[H_{s,t}|Y_1=(s',t')].$$

We first note that the hitting time of the transition chain is finite.
\begin{lemma}
    $\hite$ is finite with probability one.
    \label{lem:finite}
\end{lemma}

\begin{IEEEproof}
    This follows by first noting that the transition chain $Y_i$ is also irreducible, provided we do not include unsupported states such as $(s_3,s_1)$ or $(s_1,s_2)$. By unsupported, we mean that the transition chain, with probability 1, never visits that state, irrespective of where the $Z_i$ chain is initiated. Next, noting that support of $Y_i$ is also finite, it is a standard result of Markov chain theory that all the states of $Y_i$ are recurrent and hence $\hite$ is finite with probability one --- see for example Theorem~6.4.4 of \cite{Durrett}.
\end{IEEEproof}

Next, we state below the required concentration inequality.
\begin{lemma}
    Let $Z_i,i\geq 1$ be a Markov chain on a finite state space $S$, with an irreducible transition probability matrix. Let $g:S\times S\to[a,b]$ be any function. Then, for any initial distribution of the chain, and for any $t>0$ and any $n\geq 1$, we have
    $$
    \pr\biggl(\biggl| \sum_{i=1}^n\biggl( g(Z_{i+1},Z_i)-\ex[g(Z_{i+1},Z_i)]\biggr)\biggr| \geq t\biggr)\leq 2e^{-\frac{t^2}{2\nu^2}},
    $$
    where $\nu^2=\frac{1}{4}n(b-a)^2\hite^2$.
    \label{lem:conc}
\end{lemma}

\begin{IEEEproof}
    See Theorem~2 of \cite{moulos20}.
\end{IEEEproof}

We are now ready to prove Theorem~\ref{th:choice}.

\begin{IEEEproof}[Proof of Theorem~\ref{th:choice}]

For the choice $k>\frac{2}{(1-\epsilon)^2}-1$, we have
\begin{gather}
    1-p = (1-\epsilon)^2 > \frac{2}{k+1}. \label{eq:proofchoice:1}
\end{gather}
Hence, by Lemma~\ref{lem:expectation}, we have
\begin{align}
    \E\biggl[\sum_{i=1}^{kn_0}R_i\big| Z_i=s_1\biggr] &= \frac{2kn_0 (1-p^2) + p(3-p)(1-(-p)^{kn_0})}{(1+p)^2}\notag \\
    &> 2kn_0 \frac{1-p^2}{(1+p)^2} = 2kn_0 \frac{1-p}{1+p},\label{eq:proofchoice:2}
\end{align}
where the inequality follows by \eqref{eq:proofchoice:1}. Now, choose \(t = 2kn_0 \frac{1-p}{1+p} - 2n_0\), and note that $t>0$ using \eqref{eq:proofchoice:2}. Then, we have
\begin{align*}
    \pe(\tpi) & \stackrel{(a)}{\leq} \pr\biggl(\sum_{i=1}^{kn_0}R_i < 2 n_0\biggl|Z_1=s_1\biggr) \\
    & \leq \pr\biggl(\biggl|\sum_{i=1}^{kn_0}R_i - E[\sum_{i=1}^{kn_0}R_i|Z_1=s_1]\biggr|> E[\sum_{i=1}^{kn_0}R_i|Z_1=s_1] - 2n_0\biggl|Z_1=s_1\biggr) \\
    & \stackrel{(b)}{\leq} \pr\biggl(\biggl|\sum_{i=1}^{kn_0}R_i - E[\sum_{i=1}^{kn_0}R_i|Z_1=s_1]\biggr|>t\biggl|Z_1=s_1\biggr) \\ 
    & \stackrel{(c)}{\leq} 2e^{-\frac{t^2}{2k n_0 \hite^2}} \\
    &= 2e^{-\frac{2kn_0}{\hite^2}(\frac{1-p}{1+p} - \frac{1}{k})^2},
\end{align*}
where $(a)$ uses Theorem~\ref{th:correct}, $(b)$ follows by \eqref{eq:proofchoice:2} and the choice of $t$, and $(c)$ uses the concentration inequality in Lemma~\ref{lem:conc} with the choice of the initial distribution being $\pr(Z_1=s_1)=1$, and by noting that $R_i\in[0,2]$. Theorem~\ref{th:choice} then follows by Lemma~\ref{lem:finite} and \eqref{eq:proofchoice:1}.
    
\end{IEEEproof}

We now conclude this section by using Theorem~\ref{th:choice} to prove Theorem~\ref{th:strong}.

\begin{IEEEproof}{Proof of Theorem~\ref{th:strong}}
    Since, Algorithm~\ref{alg:sim} consists of $k$ rounds, and each round is a pair of transmissions, it has a rate of $\frac{1}{2k}$. Now, by Theorem~\ref{th:choice}, we note that the simulation in Algorithm~\ref{alg:sim} has its probability of error $\pe(\tpi)\to 0$ if $k>\frac{2}{(1-\epsilon)^2}-1$.  This implies that any rate below $\frac{(1-\epsilon)^2}{2(2-(1-\epsilon)^2)}$ is achievable. Since, $\capint(\epsilon)$ is the supremum of all achievable rates, we must have 
$$
\capint(\epsilon)\geq \frac{(1-\epsilon)^2}{2(2-(1-\epsilon)^2)}
$$
is achievable. This completes the proof of Theorem~\ref{th:strong}.
\end{IEEEproof}

\section{Channel reduction and proof of Theorem~\ref{th:main}}\label{sec:repeat}

Firstly, note that the Shannon capacity of $\BEC(\epsilon)$ is simply $\capsh(\epsilon)=1-\epsilon$ --- see for example Section~7.1.5 of \cite{CT06}. We first show using Theorem~\ref{th:strong} that for any $\epsilon<0.61484$ Theorem~\ref{th:main} holds. Since \((2-(1-\epsilon)^2) > 0\), the statement is equivalent to finding the values of \(\epsilon\) where \((1-\epsilon)^2 + \frac{(1-\epsilon)}{0.208}-2 > 0\). For such a choice of $\epsilon$, we must have
\begin{align*}
     &\epsilon^2 - 6.80769 \epsilon + 3.80769 > 0 \\
    \implies &(\epsilon - 6.19284)(\epsilon - 0.61484) > 0 
\end{align*}
Since $\epsilon<1$, we require \(\epsilon < 0.61484\) for the statement to be true. It therefore remains to show Theorem~\ref{th:main} for $\epsilon\geq 0.61484$.

The key idea to prove Theorem~\ref{th:main} for $\epsilon\geq 0.61484$ is using a repetition code to reduce the erasure probability, and then run Algorithm~\ref{alg:sim} on this reduced channel. We will first require the technical lemma which shows what rates are achievable using this repetition scheme. 

\begin{lemma}
    If a rate \(r\) is achievable for \(\BEC(\epsilon')\) for a fixed \(0 < \epsilon' < 1\), then for all $\epsilon \in(\epsilon', 1)$ the following holds:
    \begin{align*}
        \capint(\epsilon) > \frac{r}{1 - \ln (\epsilon')}\capsh(\epsilon).
    \end{align*}
    \label{lem:repeat}
\end{lemma}

\begin{IEEEproof}
For any \(\epsilon' < \epsilon \leq 1\), given a channel with erasure probability \(\epsilon\), we can reduce the channel to one with erasure probability at most \(\epsilon'\) by repeating our transmissions \(\rho\) times. To do this, we require a \(\rho\) such that \(\epsilon^\rho \leq \epsilon'\). Hence, we require
\begin{gather*}
    \rho \ln \epsilon \leq \ln \epsilon'
    \implies \rho \geq \left \lceil \frac{\ln \epsilon'}{\ln \epsilon} \right \rceil. 
\end{gather*}
Therefore, we choose $\rho=\lceil\frac{\ln \epsilon'}{\ln \epsilon}\rceil$, and hence, $\rho \leq 1 + \frac{\ln \epsilon'}{\ln \epsilon}$.

Now, using \(\rho\)-repetition on any coding scheme for the BEC(\(\epsilon'\)) gives a valid coding scheme for BEC(\(\epsilon\)). Hence \(\capint(\epsilon) \geq \frac{\capint(\epsilon')}{\rho}\). Therefore, we have
\begin{align*}
    \frac{\capint(\epsilon)}{1 - \epsilon} & \geq \frac{\capint(\epsilon')}{\rho(1-\epsilon)}\\
    &\stackrel{(a)}{\geq} \frac{\capint(\epsilon')}{(1-\epsilon) + (1-\epsilon)\frac{\ln \epsilon'}{\ln \epsilon}} \\
    & \stackrel{(b)}{\geq} \frac{\capint(\epsilon')}{(1-\epsilon) - \ln(\epsilon)\frac{\ln \epsilon'}{\ln \epsilon}} \\
    & \geq \frac{\capint(\epsilon')}{1-\ln \epsilon'}, 
\end{align*}
where $(a)$ follows using the choice for $\rho$ and $(b)$ uses the fact that \(1-x \leq -\ln(x)\) for \(0 \leq x \leq 1\). Lemma~\ref{lem:repeat} then follows by noting $\capsh(\epsilon)=1-\epsilon$.
\end{IEEEproof}

Finally, observe that the choice $\epsilon'=0.073$ yields $\capint(\epsilon')\geq 0.377$ using Theorem~\ref{th:strong}. Using this choice of $\epsilon'$ in Lemma~\ref{lem:repeat}, Theorem~\ref{th:main} holds for all $\epsilon>0.073$. This completes the proof of Theorem~\ref{th:main}.

\section{Conclusion}\label{sec:conc}

In this paper, we improved the lower bound on the interactive capacity of binary erasure channel by a factor of roughly 1.75 over the best existing bound of \cite{BKOS21}. The improvement is due to two reasons. Firstly, we tailored our simulation specifically to the erasure channel, as opposed to the more general result of \cite{BKOS21} applicable to all binary memoryless symmetric channels. The key improvement however comes from our technique of error pattern analysis, which specifically bounds the probability of occurrence of harmful error patterns, as opposed to bounding the probability of the minimum number of errors needed to corrupt a simulation. 

We also note an important drawback of our result. It is obvious that in the absence of noise, i.e., $\epsilon=0$, we must have $\capint(\epsilon)=1$. However, our bound in Theorem~\ref{th:strong} only evaluates to $\frac{1}{2}$ at $\epsilon=0$. We remark that this weakness is not due to the lack of tightness in analysis, but rather because our simulation protocol in Algorithm~\ref{alg:sim} runs on a two-bit alphabet. We leave the task of mitigating this issue as future work.

\bibliographystyle{IEEEtran}
\bibliography{references}

\begin{thebibliography}{10}
\providecommand{\url}[1]{#1}
\csname url@samestyle\endcsname
\providecommand{\newblock}{\relax}
\providecommand{\bibinfo}[2]{#2}
\providecommand{\BIBentrySTDinterwordspacing}{\spaceskip=0pt\relax}
\providecommand{\BIBentryALTinterwordstretchfactor}{4}
\providecommand{\BIBentryALTinterwordspacing}{\spaceskip=\fontdimen2\font plus
\BIBentryALTinterwordstretchfactor\fontdimen3\font minus \fontdimen4\font\relax}
\providecommand{\BIBforeignlanguage}[2]{{%
\expandafter\ifx\csname l@#1\endcsname\relax
\typeout{** WARNING: IEEEtran.bst: No hyphenation pattern has been}%
\typeout{** loaded for the language `#1'. Using the pattern for}%
\typeout{** the default language instead.}%
\else
\language=\csname l@#1\endcsname
\fi
#2}}
\providecommand{\BIBdecl}{\relax}
\BIBdecl

\bibitem{S96}
L.~J. {Schulman}, ``Coding for interactive communication,'' \emph{IEEE Transactions on Information Theory}, vol.~42, no.~6, pp. 1745--1756, 1996.

\bibitem{KR13}
\BIBentryALTinterwordspacing
G.~Kol and R.~Raz, ``Interactive channel capacity,'' in \emph{Proceedings of the Forty-Fifth Annual ACM Symposium on Theory of Computing}, ser. STOC '13.\hskip 1em plus 0.5em minus 0.4em\relax New York, NY, USA: Association for Computing Machinery, 2013, p. 715–724. [Online]. Available: \url{https://doi.org/10.1145/2488608.2488699}
\BIBentrySTDinterwordspacing

\bibitem{CT06}
T.~M. Cover and J.~A. Thomas, \emph{Elements of Information Theory (Wiley Series in Telecommunications and Signal Processing)}.\hskip 1em plus 0.5em minus 0.4em\relax USA: Wiley-Interscience, 2006.

\bibitem{BKOS21}
A.~Ben-Yishai, Y.-H. Kim, O.~Ordentlich, and O.~Shayevitz, ``A lower bound on the essential interactive capacity of binary memoryless symmetric channels,'' \emph{IEEE Transactions on Information Theory}, vol.~67, no.~12, pp. 7639--7658, 2021.

\bibitem{EGH16}
K.~Efremenko, R.~Gelles, and B.~Haeupler, ``Maximal noise in interactive communication over erasure channels and channels with feedback,'' \emph{IEEE Transactions on Information Theory}, vol.~62, no.~8, pp. 4575--4588, 2016.

\bibitem{RS94}
\BIBentryALTinterwordspacing
S.~Rajagopalan and L.~Schulman, ``A coding theorem for distributed computation,'' in \emph{Proceedings of the Twenty-Sixth Annual ACM Symposium on Theory of Computing}, ser. STOC '94.\hskip 1em plus 0.5em minus 0.4em\relax New York, NY, USA: Association for Computing Machinery, 1994, p. 790–799. [Online]. Available: \url{https://doi.org/10.1145/195058.195462}
\BIBentrySTDinterwordspacing

\bibitem{howard}
R.~Howard, \emph{Dynamic programming and Markov processes}.\hskip 1em plus 0.5em minus 0.4em\relax Technology Press of Massachusetts Institute of Technology, 1960.

\bibitem{bona}
M.~Bona, \emph{Walk Through Combinatorics, A: An Introduction To Enumeration And Graph Theory (Fourth Edition)}.\hskip 1em plus 0.5em minus 0.4em\relax World Scientific Publishing Company, 2016.

\bibitem{moulos20}
V.~Moulos, ``A hoeffding inequality for finite state markov chains and its applications to markovian bandits,'' in \emph{2020 IEEE International Symposium on Information Theory (ISIT)}, 2020, pp. 2777--2782.

\bibitem{gelles17}
R.~Gelles, ``Coding for interactive communication: A survey,'' \emph{Foundations and Trends® in Theoretical Computer Science}, vol.~13, no. 1–2, pp. 1--157, 2017.

\bibitem{BR14}
M.~Braverman and A.~Rao, ``Toward coding for maximum errors in interactive communication,'' \emph{IEEE Transactions on Information Theory}, vol.~60, no.~11, pp. 7248--7255, 2014.

\bibitem{GZ23}
\BIBentryALTinterwordspacing
M.~Gupta and R.~Y. Zhang, ``Efficient interactive coding achieving optimal error resilience over the binary channel,'' in \emph{Proceedings of the 55th Annual {ACM} Symposium on Theory of Computing, {STOC} 2023, Orlando, FL, USA, June 20-23, 2023}, B.~Saha and R.~A. Servedio, Eds.\hskip 1em plus 0.5em minus 0.4em\relax {ACM}, 2023, pp. 1449--1462. [Online]. Available: \url{https://doi.org/10.1145/3564246.3585162}
\BIBentrySTDinterwordspacing

\bibitem{GH17}
R.~Gelles and B.~Haeupler, ``Capacity of interactive communication over erasure channels and channels with feedback,'' \emph{SIAM Journal on Computing}, vol.~46, no.~4, pp. 1449--1472, 2017.

\bibitem{JSAIT}
M.~Mukherjee and R.~Gelles, ``Multiparty interactive coding over networks of intersecting broadcast links,'' \emph{IEEE Journal on Selected Areas in Information Theory}, vol.~2, no.~4, pp. 1078--1092, 2021.

\bibitem{EKS18}
K.~Efremenko, G.~Kol, and R.~Saxena, ``Interactive coding over the noisy broadcast channel,'' in \emph{Proceedings of the 50th Annual {ACM} {SIGACT} Symposium on Theory of Computing, {STOC} 2018}, 2018, pp. 507--520.

\bibitem{EKPS21}
K.~Efremenko, G.~Kol, D.~Paramonov, and R.~R. Saxena, ``Tight bounds for general computation in noisy broadcast networks,'' in \emph{2021 IEEE 62nd Annual Symposium on Foundations of Computer Science (FOCS)}, 2022, pp. 634--645.

\bibitem{clique}
\BIBentryALTinterwordspacing
N.~Alon, M.~Braverman, K.~Efremenko, R.~Gelles, and B.~Haeupler, ``Reliable communication over highly connected noisy networks,'' in \emph{Proceedings of the 2016 ACM Symposium on Principles of Distributed Computing}, ser. PODC '16.\hskip 1em plus 0.5em minus 0.4em\relax New York, NY, USA: Association for Computing Machinery, 2016, p. 165–173. [Online]. Available: \url{https://doi.org/10.1145/2933057.2933085}
\BIBentrySTDinterwordspacing

\bibitem{cycle}
R.~{Gelles} and Y.~T. {Kalai}, ``Constant-rate interactive coding is impossible, even in constant-degree networks,'' \emph{IEEE Transactions on Information Theory}, vol.~65, no.~6, pp. 3812--3829, 2019.

\bibitem{Star}
\BIBentryALTinterwordspacing
M.~Braverman, K.~Efremenko, R.~Gelles, and B.~Haeupler, ``Constant-rate coding for multiparty interactive communication is impossible,'' \emph{J. ACM}, vol.~65, no.~1, Dec. 2017. [Online]. Available: \url{https://doi.org/10.1145/3050218}
\BIBentrySTDinterwordspacing

\bibitem{JKL15}
A.~Jain, Y.~T. Kalai, and A.~Lewko, ``Interactive coding for multiparty protocols,'' in \emph{Proceedings of the 6th Conference on Innovations in Theoretical Computer Science, {ITCS~'15}}, 2015, pp. 1--10.

\bibitem{HS16}
W.~M. Hoza and L.~J. Schulman, ``The adversarial noise threshold for distributed protocols,'' in \emph{Proceedings of the Twenty-Seventh Annual ACM-SIAM Symposium on Discrete Algorithms}, 2016, pp. 240--258.

\bibitem{GKR21}
R.~Gelles, Y.~T. Kalai, and G.~Ramnarayan, ``Efficient multiparty interactive coding—part~{I}: Oblivious insertions, deletions and substitutions,'' \emph{IEEE Transactions on Information Theory}, vol.~67, no.~6, pp. 3411--3437, 2021.

\bibitem{GKR22}
------, ``Efficient multiparty interactive coding—part~{II}: Non-oblivious noise,'' \emph{IEEE Transactions on Information Theory}, vol.~68, no.~7, pp. 4723--4749, 2022.

\bibitem{EKSbeep}
\BIBentryALTinterwordspacing
K.~Efremenko, G.~Kol, and R.~R. Saxena, ``Noisy beeps,'' in \emph{Proceedings of the 39th Symposium on Principles of Distributed Computing}, ser. PODC '20.\hskip 1em plus 0.5em minus 0.4em\relax New York, NY, USA: Association for Computing Machinery, 2020, p. 418–427. [Online]. Available: \url{https://doi.org/10.1145/3382734.3404501}
\BIBentrySTDinterwordspacing

\bibitem{AGL22}
\BIBentryALTinterwordspacing
Y.~Ashkenazi, R.~Gelles, and A.~Leshem, ``Noisy beeping networks,'' \emph{Information and Computation}, vol. 289, p. 104925, 2022. [Online]. Available: \url{https://www.sciencedirect.com/science/article/pii/S0890540122000785}
\BIBentrySTDinterwordspacing

\bibitem{GHKRW}
R.~Gelles, B.~Haeupler, G.~Kol, N.~Ron-Zewi, and A.~Wigderson, ``Towards optimal deterministic coding for interactive communication,'' in \emph{Proceedings of the Twenty-Seventh Annual ACM-SIAM Symposium on Discrete Algorithms}, ser. SODA '16, 2016, p. 1922–1936.

\bibitem{Haeupler14}
B.~Haeupler, ``Interactive channel capacity revisited,'' in \emph{2014 IEEE 55th Annual Symposium on Foundations of Computer Science}, 2014, pp. 226--235.

\bibitem{CS20}
G.~Cohen and S.~Samocha, ``Palette-alternating tree codes,'' in \emph{Proceedings of the 35th Computational Complexity Conference}, ser. CCC '20, 2020.

\bibitem{Durrett}
R.~Durrett, \emph{Probability: Theory and Examples}, 5th~ed., ser. Cambridge Series in Statistical and Probabilistic Mathematics.\hskip 1em plus 0.5em minus 0.4em\relax Cambridge University Press, 2019.

\end{thebibliography}

\end{document}